\documentstyle[12pt]{article}
\newcommand{\nn}{\noindent}
\newcommand{\no}{\nonumber\\}
\newcommand{\be}{\begin{equation}}
\newcommand{\ee}{\end{equation}}
\newcommand{\ba}{\begin{eqnarray}}
\newcommand{\ea}{\end{eqnarray}}
\newcommand{\ci}[1]{\cite{#1}}
\newcommand{\bi}[1]{\bibitem{#1}}
\newcommand{\la}[1]{\label{#1}}

\def\slash#1{\not\!\!#1}
\def\gl#1{(\ref{#1})}
\def\L#1{{\cal L}_{#1}}
\def\tr#1{\mbox{\rm tr}\left[#1\right]}
\def\Tr#1{\mbox{\rm Tr}\left[#1\right]}
\def\D#1{ {\cal D} #1 \,}

\newcommand{\lam}{\Lambda_{\mbox{\scriptsize C}}}
\newcommand{\Evac}{\mbox{\Large $\varepsilon_{\mbox{\scriptsize vac}}$}}
\newcommand{\Ezero}{\mbox{\Large $\varepsilon_{\mbox{\scriptsize
vac}}^{\mbox{\scriptsize (0)}}$}}
\newcommand{\Eone}{\mbox{\Large $\varepsilon_{\mbox{\scriptsize
vac}}^{\mbox{\scriptsize (1)}}$}}
\newcommand{\nf}{ \right|_{N_F=0}}
\newcommand{\bo}{\bar b_0}
\newcommand{\cond}{\langle (G^a_{\mu\nu})^2 \rangle_{n.p.}}
\date{}

\begin{document}

%%%%%%%%%%% Title Page For DESY Preprint %%%%%%%%%%%%%%%%%%%%%%%%%%%%%%%%

%%%%%%% First Page (Cover) %%%%%%%%%%%%%%%%%%%%%%%%%%%

\thispagestyle{empty}
\begin{titlepage}

\vskip5em

\vspace{1.5cm}
\begin{center}
{\baselineskip1.5em
\Large\bf Dimensional structural constants \\[\medskipamount] from chiral
and conformal bosonization of QCD}

\vspace{3em}

{\bf A. A. Andrianov},
{\bf D.\ Ebert}, 
{\bf T.\ Feldmann}, 
\\[0.2em]
{\sl Institute of Physics, Humboldt--University Berlin,\\
Invalidenstra{\ss}e 110, D--10115 Berlin, Germany}
\\[1em]
{\bf V. A. Andrianov}, 
\\[0.2em]
{\sl Department of Theoretical Physics,
St.-Petersburg State University,\\
198904 St.-Petersburg, Russia,}
\end{center}

\vspace{3cm}

\vfill
\end{titlepage}

%%%%%%%%%%%%%%% second page (title, abstract, footnotes) %%%%
\newpage
\thispagestyle{empty}
\begin{titlepage}

\begin{flushright}
SPbU--IP--96--34\\
DESY 96--267\\
HUB--EP--96/64 \\
hep-ph/9702338 
\end{flushright}

\vspace{1.5cm}
\begin{center}
{\baselineskip3em
\Large\bf Dimensional structural constants \\[\medskipamount] from chiral
and conformal bosonization of QCD}

\vspace{1cm}

\renewcommand{\thefootnote}{\fnsymbol{footnote}}

{\bf A. A. Andrianov}\footnotemark[1]\footnotemark[3]\footnotemark[6],
{\bf D.\ Ebert}\footnotemark[2] 
and {\bf T.\ Feldmann}\footnotemark[4], 
\\[0.2em]
{\sl Institute of Physics, Humboldt--University Berlin,\\
Invalidenstra{\ss}e 110, D--10115 Berlin, Germany}
\\[1em]
{\bf V. A. Andrianov}\footnotemark[3], 
\\[0.2em]
{\sl Department of Theoretical Physics,
St.-Petersburg State University,\\
198904 St.-Petersburg, Russia,}
\end{center}

\vspace{3cm}

\begin{abstract}
We derive the dimensional non-perturbative part of the QCD effective
action for scalar and pseudoscalar meson fields by means of
chiral and conformal bosonization. The related structural
coupling constants $L_5$ and $L_8$ of 
the chiral lagrangian are estimated using general relations which
are valid in a variety of
chiral bosonization models without explicit reference to
model parameters.
The asymptotics for large scalar fields in QCD is elaborated,
and model-independent constraints on
dimensional coupling constants of the effective
meson lagrangian are evaluated.  
We determine also the interaction 
between  scalar quarkonium and the gluon density 
and obtain the scalar glueball-quarkonium potential.
\end{abstract}

\vfill

\nn
------------------------------------- \\

\setcounter{footnote}{1}
\renewcommand{\thefootnote}{\fnsymbol{footnote}}
\footnotetext{
Supported by
Deutsche Forschungsgemeinschaft under contract 436~RUS~113-29.}%\\
\setcounter{footnote}{2}
\footnotetext{
Supported by
Deutsche Forschungsgemeinschaft under contract Eb 139/1-2.}%\\
\setcounter{footnote}{3}
\footnotetext{
Supported by  RFBR (Grant No. 95-02-05346a),
by  INTAS (Grant No. 93-283ext) and by GRACENAS
(Grant No. 95-0-6.3-13).}%\\
\setcounter{footnote}{4}
\footnotetext{
Supported by
the Graduiertenkolleg {\it {\lq}Structure, Precision Tests and Extensions
                       of the Standard Model of Elementary
                       Particle Physics{\rq}}.}
\setcounter{footnote}{6}
\footnotetext{On leave of absence from
{\sl Department of Theoretical Physics,
St.-Petersburg State University,
198904 St.-Petersburg, Russia,}}
\end{titlepage}

%%%%%%%%%%%%%%%%% end titlepage desy %%%%%%%%%%%%%%%%%%%%%%%%%%%%%%%%%%

%%%%%%%%%%%%%%%%%%%%%%%%%%%%%%%%%%%%%%%%%%%%%%%%%%%%%
% more space between lines
 \baselineskip1.5em
%%%%%%%%%%%%%%%%%%%%%%%%%%%%%%%%%%%%%%%%%%%%%%%%%%%%%
\renewcommand{\thefootnote}{\arabic{footnote}}
\setcounter{footnote}{0}
\setcounter{page}{1}
%%%%%%%%%%%%%%%%%%%%%%%%%%%%%%%%%%%%%%%%%%%

%%%%%%%%%%%%%%%%%%%%%%% introduction %%%%%%%%%%%%%%%%%%%%%%%%
%%%%%%%%%%%%%%%%%%%%%%%%%%%%%%%%%%%%%%%%%%%%%%%%%%%%%%%%%%%%%%%%%%%%%%%%

\section{Introduction}
\hspace*{3ex}
 The low-energy effective action for  light
mesons \ci{We} describes their
strong and electroweak interactions by means of several 
structural constants \ci{GL} which contain the information about 
the dynamics of quark and gluon interaction  -- Quantum 
Chromodynamics (QCD).  These constants can be evaluated from the 
analysis of experimental data \ci{GL,Do}, thereby giving a 
possibility to examine QCD at low energies.  

For the 
calculation of coupling constants of the chiral lagrangian 
direct bosonization methods of low-energy  QCD
%\ci{AA,KS,Sim,DP,AANN,MM,DN,Fr} 
\ci{AA}--\ci{Fr}
and 
quark models 
%\ci{EV,Dh,Zuk,ER,Vo,ERT,EbB,BBR} 
\ci{EV}--\ci{BBR}
have been developed (see reviews 
\ci{Vo,Bi}).  These methods give good and stable numerical 
estimates for certain dimensionless chiral coefficients, 
namely for $L_1, L_2, L_3, L_4, L_6, L_9, L_{10}$ 
\ci{AA},\cite{ER}--\cite{EbB} 
in the $SU_F(3)$ case and for 
$L_{16}, L_{17}, L_{19}$ \ci{AAY} in the $U_A(1)$ extension \ci{DW} of 
the chiral lagrangian.  Meantime the chiral coefficients
$L_5, L_7, L_8, L_{14}, L_{15}, L_{18}$
depending explicitly on QCD order parameters (such as the 
quark and gluon condensate \ci{No}) were not so far calculated 
in a model-independent way (see however large-$N_c$ 
relations between them in \ci{AAY}).
Thus, it is one of the goals of our paper, to find new 
constraints  on those constants, in particular $L_5$ and $L_8$, 
from basic properties of the QCD vacuum \ci{AVA}. We analyze the common 
structure of effective lagrangians for pseudoscalar and scalar
mesons as they are derived in general by the low-energy QCD bosonization
procedure,
and impose on such lagrangians the conditions to reproduce
the behavior of the non-perturbative  QCD vacuum energy in the 
large mass limit. These conditions together with a minimal number of
physical inputs allow us to find the dimensional chiral 
coefficients, in particular $L_5$ and $L_8$.
In Chiral Pertubation Theory (ChPT)  the constant $L_5$ enters
the determination of the ratio of weak decay constants 
$F_\pi : F_K : F_\eta$ of pseudoscalar mesons.
The constant $L_8$ characterizes the $K^0-K^+$ meson mass difference
as well as the current quark mass ratio
$(2 m_s - m_d -m_u ) : (m_d-m_u)
$ \ci{GL}.

In a general QCD bosonization approach 
the dynamical chiral 
symmetry breaking (DCSB) is implemented in order to
derive the effective meson lagrangian. The DCSB is 
typically simulated by means of a 
quark momentum cutoff $\Lambda $ and a quark 
spectrum asymmetry $M$ in QCD \ci{AA,AANN} or a 
dynamical quark mass $M_d$ in chiral quark models \ci{EV,DP,ERT} 
and quark models of Nambu-Jona-Lasinio type
%\ci{ER,Vo,ERT,EbB,BBR}
\ci{ER}--\ci{Vo}, \ci{EbB}, \ci{BBR}.  The 
low-energy meson lagrangian is obtained from the bosonization 
of the quark determinant  by application of the derivative expansion
combined with the expansion in inverse powers of $M$ 
or $M_d$. It contains functionals of meson fields and external
sources of different canonical dimension with coupling 
constants proportional to  the DCSB order parameters.
We find that the very structure of the above-mentioned functionals
is universal  in respect to different bosonization schemes and the  
model-dependent information is collected in structural
constants.
Just this observation together with QCD vacuum energy 
constraints entails the definite predictions on chiral 
coefficients. As an interesting result, in 
addition we derive the scalar 
glueball-quarkonium potential directly from the QCD
generating functional.

The starting point of our analysis
is the conventional QCD generating functional for
colorless quark and gluon currents,
\ba
&&Z(\rho,V,A,S,P)  \equiv \no
&&\int \D{G} \D{q} \D{\bar q} \,
\exp \left\{
- \int d^4 x \left[
\left( \frac{1}{g^2} + \rho \right) \, \frac{1}{4} (G_{\mu\nu}^a)^2
 + \bar q (\slash D  + i S + \gamma_5 P ) q \right] \right\} \ .
\label{eq1} 
\ea
Throughout, we will work in the Euclidean space. The covariant
derivative is given by
$\slash D = i \gamma^\mu (\partial_\mu - i G_\mu^a \lambda^a 
+ V_{\mu} + \gamma_5 A_{\mu})$
with
$\tr{\lambda^a \lambda^b} = 1/2\, \delta^{ab}$ in color 
space. The external isoscalar and isovector colorless sources
$\rho(x), V_{\mu}(x), A_{\mu}(x), S(x), P(x)$ serve for 
the building of Green functions of colorless currents as usual.
The gauge fixing and ghost terms are not quoted 
explicitly but are understood to be taken into account in 
perturbative calculations.

The derivation of the scalar-pseudoscalar meson lagrangian from 
QCD by direct bosonization \ci{AA}--\cite{Fr} or by bosonization 
of a quark model \ci{EV}--\cite{BBR} can be described 
schematically as follows, 
\ba
Z(\rho,V,A,S,P) 
&=& \int\!{\cal D}G\,{\cal D}q{\cal D}\bar q \D{h}
\delta(h - (G_{\mu\nu}^a)^2) 
 \exp (- S(\bar q,\,q,\,G;\,\rho,\,V,\,A,\,S,\,P))\no
&\simeq& \int\! 
\D{h} \D{\Sigma} \D{U} \exp(- S_{ eff}( h, 
\Sigma, U;\,\,\rho, V, A, S, P) 
%\cdot {\cal R} 
\ ,  \la{boso} 
\ea 
where the averaging over gluons $G_{\mu}$ and quarks $q, 
\bar q$ is approximately replaced by the averaging  over chiral fields $U$
as well as over 
scalar glueball $h$ and quarkonium $\Sigma$ fields. 
Next the effective action is  
expanded in derivatives of meson fields and external sources.  
The model design consists  in the choice of collective
bosonic fields describing light mesons and of non-perturbative parameters 
specifying 
the DCSB (the type and magnitude of a momentum cutoff, the spectral 
asymmetry or the dynamical mass).
% and in the assumption of best fit
%that the remainder ${\cal R} \simeq 1$ at low energies.

The paper is organized as follows: 
In the next section we discuss the relations
between the structure constants of the phenomenological
low-energy meson lagrangian with those of 
model calculations. We will show that the
values of dimensional structural constants, especially, of 
$L_5$ and $L_8$ are rather model-dependent.
In section~3 we therefore extend the QCD bosonization
approach to the scalar sector including all 
possible dimensional coupling constants.
Comparing with the large scalar field asymptotics of
the QCD vacuum energy, which is derived in section~4,
we find additional constraints on the chiral structure
constants which are discussed in section~5. In addition,
this yields the scalar glueball-quarkonium potential.
A summary of the results and a conclusion are given
in section~6.

\section{QCD bosonization in the pseudoscalar sector}

In the chiral field sector ($h, \Sigma \simeq const$) 
the effective lagrangian  has the
following structure,
\be
S_{ eff} = \int d^4x \,  ({\cal L}_2  + {\cal L}_4) \,+\, S_{WZW}\ ,
\la{lag1}
\ee
Here, the Weinberg lagrangian which is of chiral dimension~2 
is given by
\be
{\cal L}_2 = \frac{F^2_0}4 \tr{(D_{\mu}U)^{\dagger} D^{\mu}U
-  \left({\chi}^{\dagger}U + U^{\dagger}\chi \right)} .\la{lag2}
\ee
where $F_0 \simeq 90$~MeV is the (bare) pion-decay constant,
$U$ is the usual $SU_F(3)$-chiral field describing pseudoscalar mesons,
$U U^{\dagger} = 1$, $ \det U = 1$. 
The external sources are assembled in the covariant derivative,
$ D_{\mu}=\partial_{\mu} + [V_{\mu},*] + \{A_{\mu},*\} $ with vector sources
$ V_{\mu}$ and axial-vector sources $A_{\mu}$  and in the
complex density 
\be
\chi = 2 B_0( S + i P) \ ,% \qquad \langle S \rangle = m_q \ ,
\label{chi}
\ee
 with scalar sources $S$ and pseudoscalar
sources $P$. The current quark mass $m_q$, as usual, is
included in the scalar source.
The constant $B_0$ is related to the quark condensate,
$i \langle \bar q q\rangle = - B_0 \, F_0^2 $,  
the order parameter of DCSB in QCD and appears in the
Gell-Mann--Oakes--Renner relation for pseudoscalar meson
masses \cite{GL}.

In the large-$N_c$ bosonization approach, i.e.\  when restricting
oneself to
the calculation of diagrams with only one quark-loop\footnote{This
yields only terms with one connected trace operation.},
the leading contribution
to the effective lagrangian ${\cal L}_4$ of chiral dimension~4 
is para\-met\-ri\-zed \ci{AAY}
by nine structural constants $I_k$:
\ba
{\cal L}_4^{ eff} &=&
\tr{- I_1\,  D_{\mu}U(D_{\nu}U)^{\dagger}D_{\mu}U(D_{\nu}U)^{\dagger} -
 I_2\, D_{\mu}U(D_{\mu}U)^{\dagger}D_{\nu}U(D_{\nu}U)^{\dagger} \right.\no
&& \left. - I_3\,  (D_{\mu}^2U)^{\dagger}D_{\nu}^2U +
I_4 \,\left( (D_{\mu}\chi)^{\dagger}D_{\mu}U + D_{\mu}\chi
(D_{\mu}U)^{\dagger}\right) \right.\no
&& \left. + I_5 \, D_{\mu}U(D_{\mu}U)^{\dagger}(\chi
U^{\dagger} + U{\chi}^{\dagger})
- I_6 \,\left( U{\chi}^{\dagger}U{\chi}^{\dagger} + \chi U^{\dagger}\chi
U^{\dagger}\right) \right.\no
&& \left.  - I_7 \,\left({\chi}^{\dagger}U - U^{\dagger}\chi \right) 
\tr{\chi^{\dagger}U - U^{\dagger}\chi} \right.\no
&& \left.  - I_8\, \left( F_{\mu \nu}^R D_{\mu}U(D_{\nu}U)^{\dagger} + F_{\mu
\nu}^L(D_{\mu}U)^{\dagger}D_{\nu}U\right) - I_9\, U^{\dagger}F_{\mu
\nu}^R U F^{L}_{ \mu \nu} }  \no
&&  + \L{4}^{ inv}\ ,
\la{mf}
\ea
in Euclidean denotations.  Herein 
$F_{\mu \nu}^L={\partial}_{\mu}L_{\nu}-{\partial}_{\nu}
L_{\mu}+[L_{\mu} , L_{\nu}]$ and\\
$ L_{\mu}=V_{\mu}+A_{\mu};\,  R_{\mu}=V_{\mu}-A_{\mu}$.

The constants $I_i \, (i \not= 7)$ arise from the bosonization 
of one-loop quark diagrams in the soft-momentum expansion.
The coefficient $I_7$ is essentially saturated \ci{Ve} by the non-perturbative
correlator of gluon pseudoscalar densities (see \ci{AAY} and references
therein).  The bosonization yields also vertices of dimension~4 
in $\L{4}^{ inv}$ which are 
invariant under chiral transformations of external sources,
\be
\L{4}^{ inv} = - H_1\, \tr{F_{\mu \nu}^L F_{\mu \nu}^L +
F_{\mu \nu}^R F_{\mu \nu}^R}  - H_2\, \tr{\chi^{\dagger} \chi}\ .
\ee
The Wess-Zumino-Witten action $S_{WZW}$ includes the
so-called anomalous vertices
\ci{GL,Do} and it is not displayed here.

The low-energy phenomenology of pseudoscalar mesons is described by the
standard Gasser-Leutwyler lagrangian \ci{GL} which
contains ten structural chiral constants $L_i$,
\ba
{\cal L}_4^{GL}&=&
- L_1\left(\tr{(D_{\mu}U)^{\dagger}D_{\mu}U}\right)^2 
- L_2 \, \tr{(D_{\mu}U)^{\dagger}D_{\nu}U}\!
\tr{ (D_{\mu}U)^{\dagger}D_{\nu}U} \no
&&- L_3\, \tr{(D_{\mu}U)^{\dagger}D_{\mu}U(D_{\nu}U)^{\dagger}
D_{\nu}U} +
L_4\,\tr{(D_{\mu}U)^{\dagger}D_{\mu}U} \tr{{\chi}^{\dagger}U
+ U^{\dagger}\chi} \no
&&+ L_5\,\tr{(D_{\mu}U)^{\dagger}D_{\mu}U({\chi}^{\dagger}U
+ U^{\dagger}\chi)}
 - L_6 \left(\tr{ {\chi}^{\dagger}U + U^{\dagger}\chi }\right)^2 \no
&& - L_7\left(\tr{{\chi}^{\dagger}U - U^{\dagger}\chi}\right)^2 -
L_8 \,\tr{{\chi}^{\dagger}U{\chi}^{\dagger}U +
U^{\dagger}\chi U^{\dagger}\chi }\no
&&- L_9\, \tr{ F_{\mu \nu}^R D_{\mu}U(D_{\nu}U)^{\dagger} + F_{\mu
\nu}^L(D_{\mu}U)^{\dagger}D_{\nu}U} - L_{10} \, \tr{ U^{\dagger}F_{\mu
\nu}^R U F^{L \mu \nu}} \no
 &&+ \L{4}^{inv}  \ .\la{lag3}
\ea
We will neglect meson-loop corrections in the
following\footnote{Consequently, we will not achieve higher precision
than the usual logarithmic uncertainty due to the choice of
renormalization scale.}, keeping our
attention to the leading order in $1/N_c$.
Then the (tree-level) 
relations between coefficients $L_i$ and $I_j$ are derived 
following the usual on-shell scheme 
of Chiral Perturbation Theory,  i.e.\
imposing the equations of motion from the Weinberg lagrangian ${\cal L}_2$,
\be
U^{\dagger}D_{\mu}^2U - (D_{\mu}^2U)^{\dagger}U-{\chi}^{\dagger}U
 + U^{\dagger}\chi
=\frac{1}{N_F} \tr{U^{\dagger}\chi -{\chi}^{\dagger}U}\ ,
\la{Eq}
\ee
where $N_F = 3$ is the number of light flavors.
One finds,
\ba
&&2L_1=  L_2= I_1,\quad L_3 =  I_2 + I_3 - 2I_1,
\quad L_4 = L_6 = 0,\quad L_5 = I_4 + I_5,\no
&&L_7 = I_7 -\frac{1}{6} I_4 + \frac{1}{12} I_3,
\quad L_8 =  -\frac{1}{4} I_3 + \frac{1}{2} I_4 + I_6,\quad
L_{9} = I_8,\quad L_{10} =  I_9 \  .
\la{Zw0}
\ea
Thereby the predictions of a bosonization scheme can be directly
referred to physical characteristics of pseudoscalar mesons.

The low-energy QCD bosonization yields the chiral lagrangian in
the soft momentum expansion whereas the ChPT concept is based upon the
soft momentum and light meson mass expansion 
$\dim[p^2] = \dim[m^2_{\pi}] = 2$.
Therefore chiral coefficients in the bosonized action have different status
in the
canonical and chiral  dimensional analysis. Namely,
the canonical dimension of vertices containing powers of the field
$\chi$
(see eq.~\gl{chi})
is always less than the chiral one because the canonical 
$\dim[\chi] = \dim[S] = 1$  and  the chiral 
$\dim[\chi] = \dim[m^2_{\pi}] = 2$ . The corresponding structural
constants $I_{4,5,6,7}$ or $L_{4,5,6,7,8}$ and $H_2$
 are in fact dimensional in the canonical sense that becomes evident when
to supplement them with factors of $B_0$ from $\chi$. Thus these constants
as well as the coefficients of $\L{2}$ are essentially non-perturbative
as they are proportional to powers of the basic QCD scale 
\be
\lam \simeq \mu \exp\left( - \frac{1}{b_0 g^2(\mu)} \right); \quad
b_0 = \frac{11N_c - 2N_F}{24 \pi^2}, \la{b0}
\ee
where $g^2(\mu)$ is a QCD gauge coupling constant measured at a scale $\mu$.

It happens to be that the 
dimensionless structural constants $I_{1,2,3,8,9}$ or $L_{1,2,3,9,10}$ are
not  sensitive to a DCSB model and  take universal values \ci{AA,ER}, 
%in the limit of
%zero hadronization radius\footnote{This statement
%might not be precise -- but general.} $\Lambda \rightarrow \infty$ \ci{ER}
%and zero gluon corrections \ci{ERT,BBR},
\be 4 I_1= - 2 I_2= 2 I_3= I_8= 2 I_9 = \frac{N_c}{48\pi^2}.
\la{estim}
\ee
These constants are in a satisfactory agreement with phenomenological
values obtained from the strong and electromagnetic interactions 
of pseudoscalar mesons
% in neglecting their masses (the chiral limit)
\ci{AA,ER,Vo,EbB}. 

On the other hand,  the values of other,
dimensional chiral constants
(except for\\
 $L_4 = L_6 = 0$ which are suppressed in
the leading $1/N_c$ order \ci{GL}) are strongly
sensitive to modelling the DCSB by means of a momentum cutoff $\Lambda$
and a spectrum asymmetry $M$  or a dynamical mass $M_d$.
In order to estimate dimensional constants in a model independent way
we  extend the chiral lagrangian by
introducing in the next section scalar glueball, $h$
and quarkonium, $\Sigma$ variables
so that their v.e.v.'s provide
the required constants.  
Further on, we will normalize  the corresponding lagrangian 
to reproduce the
QCD motivated expansion of the vacuum energy 
for small and large external scalar sources.  It will provide
the model-independent 
constraints on coupling constants of the joint scalar-pseudoscalar
lagrangian. Finally, the large mass reduction of scalar mesons in
the latter one
yields the required
estimates on the chiral constants $L_5, L_8$.

Let us start this derivation 
with preparing the concise form of the dimensional part
of  ${\cal L}_{ch}$,
\ba
{\cal L}^{\dim}_{ch} &=& \tr{- F_0^2 \left( \tilde A_\mu^2
 + B_0 \tilde S \right)
- 8 B_0  I_4  \,i \tilde A_\mu \tilde D^V_\mu \tilde P 
- 16 B_0 \left(I_4 + I_5\right)  \tilde A_\mu^2 \tilde S \right. \no
&& \left. - 8B_0^2 I_6   \left(\tilde S^2 - \tilde P^2\right)  
- 4 B_0^2 H_2  \left(\tilde S^2 + \tilde P^2\right) } + 
\Delta{\cal L}^{\dim}_{ch} \ , 
\la{Ldim} 
\ea
where $\tilde D^V_{\mu} \equiv \partial_{\mu } + [\tilde V_{\mu},*]$. 
Here and in what follows we omit  
the $I_7$ term as we do not discuss the $U(1)_A$ terms
in this paper.
The chiral fields $U(x)$ are encoded in rotated external sources by means
of the following chiral bosonization rules,
\ba
\tilde V_\mu + \tilde A_\mu &=& U^\dagger (V_\mu + A_\mu) U
+ U^\dagger \partial_\mu U ; \qquad
\tilde V_\mu - \tilde A_\mu = V_\mu - A_\mu ; \no
\tilde S + i \tilde P &=& U^\dagger (S + i P) ; \qquad
\tilde S - i \tilde P = (S - i P) U \ . \la{dres}
\ea
With these rules one can easily reproduce the chiral lagrangian (\ref{mf}).
The terms in \gl{Ldim} are ordered in accordance with the ChPT count
and $\Delta{\cal L}^{\dim}_{ch}$ stands for vertices of dim-6 and higher.
Meanwhile from the canonical dimensional analysis it 
follows that two more vertices, namely  
\be
\Delta{\cal L}^{\dim}_{ch} =8  B_0 I_{10} \, \tilde S^3 +  
8 B_0 I_{11}  \, \tilde S \tilde P^2 + \dots \ , \la{delta}
\ee  
have the 
comparable dim-3 and they should be retained in the extended 
scalar-pseudoscalar lagrangian (see below).

The bosonization models with DCSB introduced by an asymmetric 
cutoff $\Lambda, M$ \ci{AA} or by a dynamical mass $M_d$  
\ci{EV,ER,EbB}
yield in fact more relations 
between the above constants. Let us derive them first in the
framework of the chiral quark model \ci{EV,DP}. 
We impose that the quark loop
effective action in external fields were regularized 
(see \ci{EbB,Ball}) by means of
a momentum cutoff $\Lambda_d$ so that it is left (and right)  invariant 
under local chiral transformations of external sources  like in 
\gl{dres} (we ignore here
the P-odd Wess-Zumino-Witten  action) .  
It corresponds to the definition of chirally
invariant quarks.
The leading low energy contribution to this action 
{\it in the large cutoff  approximation}
$\Lambda_d \gg |S|, |P|,|V|,|A| $ 
is given by divergent parts of loop integrals \cite{Vo,EbB,Ball},
\ba
\Gamma_+  &\simeq& \frac{N_c}{16\pi^2} \int d^4x \, 
\tr{ -  2 \, c \, \Lambda_d^2 \ (S^2 + P^2)\right. \no
 &&\left. + \ln\frac{\Lambda_d^2}{\mu^2} \, \left\{
\left((S^2 + P^2)^2 - [S, P]^2\right)
\right.\right. \no
&& \left.\left. + \left( (D^V_{\mu} S)^2 + (D^V_{\mu} P)^2 - \{A_{\mu}, S\}^2
- \{A_{\mu}, P\}^2  - 2 i D^V_{\mu}P  \{A_{\mu}, S\}  
+ 2 i D^V_{\mu}S  \{A_{\mu}, P\} \right)\right.\right. \no
&& \left.\left. - \frac13 \left( (F_{\mu\nu}^L)^2 +  (F_{\mu\nu}^R)^2\right)
\right\} } + \cdots  \la{invar}
 \ea
It consists of four independent chiral invariants 
corresponding to the four lines of \gl{invar}.
For an $O(4)$ invariant momentum  cutoff one has 
$c = 1$ but it is regularization
dependent. 
Meanwhile, the logarithmically  divergent  coefficients are unique.

The coupling of chirally invariant quarks to
pseudoscalar fields is provided by the dynamical mass term introduced 
in such a way that 
the designed chiral invariance of 
(the P-even part of)  the fermion determinant is preserved,
\be
\left(S - i \gamma_5 P\right) \longrightarrow 
\left(S - i \gamma_5 P  + M_d \left( U^{\dagger} P_L + P_R
\right)\right) \longrightarrow  \left( \tilde S - i \gamma_5 \tilde P  + M_d\right)\  , \la{shift}
\ee
following the  notations of \gl{dres}, and as well
$V_\mu,A_\mu \to \tilde V_\mu, \tilde A_\mu$.
The equivalence is provided by the invariance under quark field rotations
 $q_L \rightarrow U q_L .$ 
The shifted effective action \gl{invar} yields 
the dimensional part \gl{Ldim} of the chiral lagrangian
${\cal L}^{\dim}_{ch}$ in terms of the parameters $\Lambda_d, M_d$,
which simulate the DCSB,  
with the following predictions for the coefficients,
\ba
 &&F_0^2 \cong \frac{N_c M_d^2}{4\pi^2} \ln\frac{\Lambda_d^2}{\mu^2};\qquad
B_0 F_0^2 \cong  \frac{N_c M_d}{4\pi^2}\left(c \Lambda_d^2 - M_d^2 
\ln\frac{\Lambda_d^2}{\mu^2}
\right) ; \qquad \mu \simeq M_d; \\
&&I_4  = I_{10} = I_{11} = \frac{F_0^2}{8 B_0 M_d};\qquad  I_5 = 0; \qquad
I_6  = - \frac{F_0^2}{16 B_0^2}; \la{Ii}\\
&&H_2 =  \frac{F_0^2}{8 B_0 M_d} - \frac{F_0^2}{8 B_0^2} = I_4 + 2 I_6 \ .  
\la{chirq}
\ea

Inspecting other bosonization
models \ci{AA}--\cite{Fr}
 we find that the  relations for $I_j$ displayed in \gl{Ii}
are also valid  in the large-cutoff approximation.  
For instance, in the QCD chiral bosonization model \ci{AA} 
the DCSB is introduced
by means of an asymmetric regulator for the quark determinant,
$
\Theta\left(\Lambda^2 - \left(\slash D - i M\right)^2\right) $ which is
non-invariant under local chiral rotations.
The pseudoscalar meson fields arise 
from local chiral rotations of the quark fields
$q(x) = \left(P_L U(x) + P_R\right) q_{inv}(x)$,  
and the chiral lagrangian is made from
 the chirally non-invariant part,
\be
 Z=\int {\cal D}U\frac{Z_q(V, A, S, P)}{Z_q(\widetilde V, \widetilde
A, \widetilde S, \widetilde P)}\langle Z_{inv}\rangle_G 
\equiv \int {\cal D}U \exp ( - S_{eff}(U; V, A, S, P)) 
\langle Z_{inv}\rangle_G,
\la{bos}
\ee
where $\langle\ldots\rangle_G$ stands for the averaging over the 
gluon vacuum.
$S_{eff}$ 
does not contain any gluon fields.
The role of gluons is reduced only to the formation of the
dimensional parameters of the theory,
$(\Lambda, M) $, on the base of the equation of stability of
the low-energy  region,  supported by the gluon condensate \ci{AA}
(see the next section).
After calculation of the chirally non-invariant part of 
the quark determinant one 
comes to the chiral lagrangian \gl{mf} with dimensional vertices collected
in \gl{Ldim}. The description of the main parameters and the 
structural constants
in the minimal version of  the model in \ci{AA}  has the following form:
\ba
 &&F_0^2 \cong \frac{N_c}{4\pi^2} \left(\Lambda^2 - M^2\right);\qquad
B_0 F_0^2 \cong  \frac{N_c M}{2\pi^2}\left(\Lambda^2 - \frac13 M^2
\right) ; \\
&&I_4  = I_{10} = I_{11} = \frac{N_c M}{16 \pi^2 B_0 };\qquad  I_5 = 0; \qquad
I_6  = - \frac{F_0^2}{16 B_0^2}; \qquad  \\
&& H_2 =  - \frac{F_0^2}{4 B_0^2} = 4 I_6 \la{chirm} \ .  
\ea
When comparing with \gl{Ii} one can see that the following relations
between
the constants $I_i$ ,
\be
I_4  = I_{10} = I_{11} \ ; \qquad
I_5 = 0 \ ; \qquad
I_6  = - \frac{F_0^2}{16 B_0^2}\ ; 
\label{genrel}
\ee
are the same in both models,
 but the analytic dependence on
model parameters and the
quantitative predictions for the coefficients are 
(of course) model-dependent. 
In particular, 
the values for $H_2$ do not coincide for any choice of parameters.
As to the other constants $F_0, B_0, I_i$, 
their values may be adjusted to the same
values in both models but for an unnatural choice of 
the constituent mass parameters:
$M_d \simeq 100$~MeV  in the chiral quark model and $M \simeq 600$~MeV 
in the chiral bosonization model \ci{AA}, respectively.
In the following, we will therefore only use the general and
model-independent relations \gl{genrel} to reduce the number
of unknown parameters in the chiral lagrangian \gl{Ldim} or
its extensions including heavy scalar fields \ci{AM} (see below). 

For illustration,
let us however
check the consistency of these predictions with the phenomenology  of
pseudoscalar mesons \gl{Zw0} for 
$M_d \simeq 250$~MeV and $B_0 \simeq 1.3$~GeV ($\Lambda_d \sim 1$~GeV):
\ba
&&L_5 =\frac{F_0^2}{8 B_0 M_d} \simeq 3 \cdot 10^{-3} 
\qquad (\mbox{cf.\
$(2.26 \pm 0.14) \cdot 10^{-3}$  from \ci{Han}}) \ ; 
\no
&&L_8 = - \frac{1}{128\pi^2} + \frac12 L_5 
- \frac{F_0^2}{16 B_0^2} \simeq  0.5 \cdot 10^{-3} \quad (\mbox{cf.\ 
$(0.9 \pm 0.4)\cdot 10^{-3}$ from \ci{Bi}}) \ ; 
\la{L8}
\ea
where $N_c = 3$ is taken.
Thus the chiral quark
model displays a satisfactory agreement with experiment. Note that whereas 
$L_5 = I_4$ rather serves as an input parameter, 
which may be tuned by a suitable choice of model parameters,
the value of
$L_8$ is well calculated from all the inputs. 
A similar situation arises in
the chiral bosonization model. 
%\ldots
%when chosing the same value for $B_0$ 
%but $M \simeq 200$~MeV.

\section{QCD bosonization in the scalar sector}

In the large $N_c$ limit the main contribution
of the scalar glueball sector to the effective quarkonium lagrangian
is provided \ci{No} by the v.e.v. of the glueball field $<h> = \cond$,
i.e.\ the gluon condensate (see \gl{boso}),
\be
C_g \equiv \frac{1}{4\pi^2} \, \cond \approx (400~\mbox{MeV})^4\ .
\la{cond}
\ee
Here {\it n.p.} means that the perturbative part
has been subtracted.  
Let us associate the scalar quarkonium fields with the radial
colorless fluctuations of light quark fields,
$q(x)  = \exp\left\{ \frac{1}{2} \Sigma(x)\right\} \, q_{inv} (x)$.
Herein $q_{inv}$ are  fields invariant under dilatations.
In general, the dilatations mix flavor numbers,
$\Sigma(x) = \sigma \mbox{\bf I} + \sigma_i \mbox{\bf T}_i$ where
{\bf T}$_i$ are generators of the $SU(N_F)$ group.
This factorization can be used in the conformal bosonization procedure
to derive the effective scalar-pseudoscalar lagrangian \ci{AANN,AM}.
We evaluate it in the soft momentum region neglecting the
kinetic terms and higher-order derivatives of (heavy) scalar meson fields.

The dimensional part of the
lagrangian for the flavor-singlet (dimensionless)
scalar field $\sigma$ then has the
general form: 
\ba
{\cal L}^{\dim}_{\sigma} &=&
\tr{\frac{\sigma}{48\pi^2} \,\cond
+ \frac14 a_4 \, e^{-4\sigma}
+ a_3 \, \tilde S \, e^{-3\sigma}
\right. \no 
&& \left. 
+ \left( a_{21} \tilde S^2 +  a_{22}\, \tilde P^2- a_{23}\, \tilde A_\mu^2\right) 
\, e^{-2\sigma} \right. \no  
&& \left. 
+  \left(a_{11}\, \tilde S^3 + a_{12}\, \tilde S \tilde P^2  
- a_{13}\,  i \tilde A_\mu \tilde D^V_\mu \tilde P - 2 a_{14} \tilde A_\mu^2 \tilde S
\right) \, e^{-\sigma} } + \ldots  \la{sdim}
\ea
%The correspondence to \gl{Ldim} for $\sigma = 0$ is evident.  
%
Let us understand this form of the lagrangian in
more detail:
The first term simply
reproduces correctly the scale anomaly of the quark determinant
\ci{Fuji}--\ci{scale2}.
The following terms are ordered according to their canonical
dimension, i.e.\ they have dimensional vertices with coupling
constants which are polynomials
in the QCD scale $\lam $,
$a_{jk} \sim \lam^j$ (the dim-4 term is a purely scalar vertex).
According to the dimensionality
of the coupling constants the vertices 
have been dilated with appropriate powers
of $\exp(-\sigma)$. 
We have again neglected  dim-4
perturbative contributions and higher dimensional vertices proportional
to inverse powers of $\lam$. 
It will be shown in the next section that such an expansion
can be made consistent with the basic properties of the QCD vacuum energy.

Let us first normalize the lagrangian such that the minimum 
of ${\cal L}^{\dim}_{\sigma}$
for {\it vanishing} external sources is 
just reached at $\sigma = 0$. It corresponds to
\be
      a_4 = \frac{1}{48\pi^2} \,\cond = \frac{C_g}{12}\ . \la{a4}
\ee
In order to find the chiral structural constants $I_{4,5,6}$
one should perform the saddle point approximation for the effective
scalar $\Sigma$-field lagrangian \ci{AM} and develop the heavy
scalar mass and ChPT expansion. Of course, when the external sources 
are present, the lagrangian \gl{sdim}
does no longer possess the minimum at $\Sigma = 0$. 
The saddle point then is rather given by
\be
\Sigma_{min} = \frac{3 a_3}{4 a_4} \tilde S + \left(\frac{a_{21}}{2 a_4}
- \frac{9a_3^2}{16a_4^2}\right)\,\tilde  S^2 + \frac{1}{2a_4}
\left(a_{22}\,\tilde P^2 -  a_{23} \tilde A^2_{\mu}\right) + \cdots \
;
\qquad 
\Sigma \equiv \Sigma_{min} + \bar\Sigma \ . \la{Sigma} 
\ee
Now, in terms of the shifted variable $\bar\Sigma$, 
the minimum occurs at $\bar\Sigma = 0$.
Let us therefore define
%\footnote{We stress that
%although the form of the lagrangian remains the same,
%the true chiral coefficients $a'$ are in general different from 
%the unprimed coefficients $a$, which are involved in the
%large scalar field asymptotics of QCD (see below).}
\be
 {\cal L}^{\dim}_\sigma (\Sigma=\Sigma_m) 
        \equiv {\cal L}^{\dim}_\sigma (\sigma=0) \Bigg|_{a_{i(j)}
        \to a'_{i(j)}} \ , 
\ee
which should be then identified with
the dimensional part of the chiral lagrangian
\gl{Ldim}, ${\cal L}^{\dim}_{\sigma}(\Sigma=\Sigma_m) = 
{\cal L}^{\dim}_{ch}$. 
Consequently, the coefficients of the chiral
lagrangian \gl{Ldim} can be matched to \gl{sdim} in the following way:
\ba
&&a'_3 = a_3 = - B_0 F_0^2\ ; \no
&&a'_{21} = a_{21} - \frac{9 a^2_3}{8a_4} = - 4B_0^2 (2I_6 + H_2);
\quad a'_{22} = a_{22} =  4B_0^2 (2I_6 - H_2);
\quad a'_{23} = a_{23} = F_0^2; \no
&&a'_{11} = a_{11} - \frac{3 a_3 a_{21}}{2 a_4} + 
\frac{45 a_3^3}{32 a_4^2} = 8B_0 I_{10}; \quad a'_{12} = a_{12} 
- \frac{3 a_3 a_{22}}{2 a_4} = 8B_0 I_{11} ;\no
&& a'_{13} = a_{13} = 8 B_0 I_4;\quad  a'_{14} = a_{14} 
- \frac{3 a_3 a_{23}}{4 a_4} = 8B_0 (I_4 + I_5)\ . \la{modif} 
\ea
Just these coupling constants should be compared with the
chiral bosonization predictions. 
%In turn the coupling constants of ${\cal L}^{\dim}_{\sigma}$ do not
%necessarily coincide to similar chiral constants.
In particular, the quark  model lagrangian \gl{invar}-\gl{chirq} 
(or the bosonization model \gl{bos} - \gl{chirm})
is induced by the choice:
\ba
&&a'_{21} = F_0^2 \left(1 - \frac{B_0}{2M_d}\right)
\quad \left[\mbox{or} \ = 
\frac32 F_0^2 \right] \ ; 
\no 
&& a'_{22} = a_{21}' - a_{23}' =
-  \frac{F_0^2 B_0}{2M_d} \quad \left[\mbox{or}\ = 
\frac12 F_0^2\right] \ ;
\no 
&& a'_{11} =  a'_{12} = a'_{13} = a'_{14} = \frac{F_0^2}{M_d} 
\quad \left[\mbox{or}\ = 
\frac{N_c M}{2\pi^2}\right] \ . \la{chm}
\ea
In the next section we will remind 
basic properties of the QCD quark vacuum energy as a function
of external scalar fields 
and further derive the additional constraints on some of the $a_{jk}$.
It will allow us to reduce the 
number of input parameters in any model building.

\section{QCD vacuum energy and large scalar fields}

We go back to  the QCD 
generating functional \gl{eq1} and consider
nearly constant scalar gluonium $\rho$ and 
quarkonium $S$ sources.
The renormalized coupling constant at a given
scale $\mu$ for $N_F$ active quarks obeys 
the following RG equation
\be 
\partial_\tau \, g^2 \,  =
\beta(g^2) \,  \simeq - b_0 \, g^4;\quad
\frac{1}{g^2(\mu)} \simeq
b_0 \, \ln\frac{\mu}{\lam} =
\frac{11 N_c - 2 N_F}{24 \pi^2} \, \ln\frac{\mu}{\lam} ; \la{RG}
\ee
in the 1-loop approximation. Herein 
$ \partial_\tau \equiv \partial / \partial \ln(\mu/\lam).$

From eq.~\gl{eq1} it is obvious, that the field $\rho$ can
be reabsorbed into the definition of the coupling constant,
i.e.\ the definition of the basic QCD scale $\lam$,
\be
\frac{1}{\tilde g^2} \equiv
\frac{1}{g^2} + \rho =
b_0\, \ln \frac{ \mu}{\lam\, e^{-\rho/b_0}} \ , \quad
\widetilde \lam  \equiv  \lam \, e^{-\rho/b_0} \ .\la{lam1}
\ee
Now let us perform an expansion of the vacuum energy
 in $N_F$, the number of (light) flavors,
\be
\Evac(\rho,S) = \Ezero(\rho) + N_F \, \Eone(\rho,S) + O(N_F^2) \ .
\ee

The leading term for $N_F=0$ is purely gluonic.
The non-perturbative part of the QCD gluonic vacuum 
energy density
$\Ezero(\rho)$ is then defined as
\be
\left. 
Z(\rho, S) \, \nf
\equiv \exp \left\{ - \Omega \, \Ezero(\rho) \right\}\ ,
\label{eq5}
\ee
with $\Omega=\int d^4x$ being the space-time volume.

From dimensional arguments we obtain that
\be
\Ezero(\rho=0) = \langle \Theta_{44} \rangle_{n.p.}
= \frac{1}{4} \langle \Theta_{\mu\mu} \rangle_{n.p.}
\sim - \lam^4 \ ,
\ee
where $ \Theta_{\mu\nu}$ stands for the energy-momentum tensor.
One therefore arrives at the  RG 
equation for the energy density $
\partial_\tau \, \Ezero(0) = - 4 \, \Ezero(0)$ which has the following
(RG invariant) solution \ci{scale1} (see eqs.\gl{eq1},\gl{RG} )
given by the trace anomaly \ci{Fuji}--\ci{scale2}:
\be
\Ezero(0)= \left . \frac{1}{16 g^4} \, \beta(g^2) \,
\cond \, \nf \simeq - \frac{\bo}{16} \, \cond \ ,
\ee
where $\bo \equiv \left. b_0 \, \nf$.
When $\rho \neq 0$ one simply has to replace $\lam$ with $\tilde\lam$
from \gl{lam1}
\be
 \Ezero(\rho) = e^{-4\rho/\bo} \, \Ezero(0)
= - \frac{\pi^2 \bo}{4} \, C_g \, e^{-4\rho/\bo} \ ,
\label{eq9}
\ee
in terms of \gl{cond}.

In order to evaluate the effective potential for the scalar
glueball field $h(x)$ we follow the bosonization
ansatz \gl{boso} which 
can be rewritten in the form of a functional
Legendre transform with respect to 
the (nearly constant) field $\tilde\rho$,
\be
\widetilde Z(h) = \exp\left\{ - \Omega V_g(h) \right\}
= \int \D{\tilde\rho} \,
\exp\left\{ - \Omega \left[ \Ezero(\tilde\rho) - \frac{1}{4} \tilde\rho\, h
\right]\right\} \ . \la{glpot}
\ee
For $\Omega \rightarrow \infty$ the saddle point configuration,
\be
\tilde\rho=\tilde\rho(h) = - \frac{\bo}{4} \, \ln \frac{h}{\cond} \ , \la{saddle}
\ee
delivers the required effective potential:
\be
V_g(h) = \Ezero(0) \, \frac{h}{h_0} \,
\left(1 - \ln \frac{h}{h_0} \right)\ ,\quad  h_0 = \cond \ ,
\ee
which coincides with the expression in \cite{scale2}.

Now we proceed to the analysis 
of the quark-loop contribution in \gl{eq1}
to the vacuum energy and its dependence on the
external scalar source $S(x)\simeq const$ (other external
sources are not shown  for the time being),
\ba
&&Z(\rho,S) \equiv \exp \left\{ - \Omega \ \Evac(\rho,S) \right\} \no 
&=&
\int \D{G}
\exp\left\{ - \int d^4 x \,
  \frac{1}{4\tilde g^2} \, (G_{\mu\nu}^a)^2
+ N_F \, \frac{1}{2} \,
  \Tr{ \ln \left( \frac{{\slash D}{}^2 + \not\!\partial S + S^2}{\mu^2}
        \right) \, R } \right\}_{n.p.} \ , \label{eq10}
\ea
where the quark fields are integrated out
leading to the quark determinant. 
Here $R = R({\slash D}{}^2, \Lambda_{UV}^2)$ is a gauge invariant
regulator.
In the following we restrict ourselves to slowly varying
sources $\not\!\partial \, S \simeq \not\!\partial \rho \simeq  0 $.

The first-order term $\Eone$
gets two contributions \ci{AVA} from both
the $\beta$-function,
\be
\partial_{N_F} \, g^{-2} \simeq
        - \frac{1}{24 \pi^2} \, \ln \frac{\mu^2}{\lam^2} \ ,
\ee
and the quark determinant.
In addition, all dimensional operators are
supplemented with scaling factors
$\exp(-\rho/\bo)$ so that the higher-dimensional 
condensates\footnote{We retain the 1-loop contributions only
and therefore neglect the anomalous dimension of $\bar q q(x)$, \\
i.e.\ of  $S(x)$.}
$C^g_{2n + 4}\sim  \\
\langle \Tr{{\slash D}{}^{2n}R}\rangle_{n.p.}
\sim \Lambda_c^{2n+4}$ are rescaled according to their canonical
$\dim =2n + 4$ and \gl{lam1}. Thereby one derives that
\be
\Eone(\rho,S) =  e^{-4\rho/\bo} \,
\left\langle
  - \frac{1}{96 \pi^2} \, \ln \frac{\mu^2}{\lam^2} \, (G_{\mu\nu}^a)^2
  - \frac{1}{2} \, \tr{ \langle x |
                \ln \left( \frac{\slash D{}^2 \,
        e^{-2\rho/\bo}+ S^2}{\mu^2} \right)
        \, R | x \rangle}
\right\rangle_{n.p.} \ .
\ee
It is a RG invariant quantity \ci{AVA},
$$
   \frac{\partial \Eone}{\partial \, \ln \mu^2}
  =
 e^{-4\rho/\bo} \,
  \left\langle - \frac{1}{96 \pi^2} \, (G_{\mu\nu}^a)^2
    + \frac{1}{2} \tr{ \langle x | R | x \rangle}
  \right\rangle_{n.p.} \ = \ 0 \ ,$$
which is provided 
by the Fujikawa's theorem \cite{Fuji},
\be 
\left\langle
    \tr{ \langle x | R | x \rangle}
  \right\rangle_{n.p.}
  = \frac{1}{48 \pi^2} \, \cond \ ,
\ee
 as long as
the regulator obeys the conditions 
$R(\slash D{}^2 = 0) =1$, $R(\slash D{}^2 \to \infty) = 0$
(in fact any momentum cut-off scheme).

Let us introduce a fermionic reference scale $\Lambda_F$ \ci{AVA} for
the Dirac operator $\slash D$ as a normalization at $S=0$ in
the following way:
\be
\left\langle \tr{\langle x | \ln \frac{\slash D^2}{\Lambda_F^2} \, R
        | x \rangle} \right\rangle_G \equiv 0 \ ,
\ee
leading to
\be
\Eone(\rho,S=0) \equiv
-  e^{-4\rho/\bo} \, \frac{1}{24} \,
     \left( \ln \frac{\Lambda_F^2 }{\lam^2}
  - \frac{2\rho}{\bo} \right) \, C_g \ .
\label{eq16}
\ee
Presumably, this (fermionic) scale is of the order
of 1~GeV, $\Lambda_F > \lam$.
Then we can redefine the quark vacuum energy using this scale,
\ba
\Eone ( \rho,S )
&=&
 \Eone ( \rho, 0) +
e^{-4\rho/\bo} \, W_F(\bar S)
\ , \no [0.15em]
W_F(\bar S) &\equiv &
- \frac{1}{2}
\, \left\langle \tr{ \langle x |
\ln \left( \frac{{\slash D}{}^2  + \bar S^2}{\Lambda_F^2}
\right) \, R | x \rangle } \right\rangle_{n.p.} \ ,
\label{eq17}
\ea
where we have defined $\bar S \equiv S \, e^{\rho/\bo}$.
When other external sources are present, they should be
rescaled in the same way, according to their canonical
dimension, $(\bar P, \bar D_\mu^V, \bar A_\mu)
= \\ (P,D^V_\mu,A_\mu) \, e^{\rho/\bo}$.

The glueball potential is again given by the Legendre transformation 
\gl{glpot}, and 
to the first order in $N_F$ 
the quark contribution is described by \gl{eq17} where
for the field $\rho$ the saddle point value 
$\rho := \tilde\rho(h)$ \gl{saddle} is taken. 
These rules can be 
straightforwardly applied to QCD bosonization models in order to derive the joint
scalar glueball-quarkonium potential,  
\be
\exp\left[- \Omega \, W_F(\bar S) \, e^{-4\tilde\rho(h)/\bo}\right]
\simeq
\int {\cal D}\sigma \,
\exp\left[- \Omega \, \left({\cal L}_\sigma (\bar S) - \frac{N_F \,
C_g}{48} \right) \,  
        e^{-4\tilde\rho(h)/\bo}\right] \ .
\label{wftosdim}
\ee
The constant in the exponential on the r.h.s.\ is fixed by the
normalization $W_F(\bar S = 0) = 0$ provided at the saddle point
$\sigma = 0$, and for our purposes ${\cal L}_\sigma$
is taken from eq.~\gl{sdim}.
After substituting \gl{wftosdim} into
the QCD generating functional \gl{eq10} and performing the
shift $\sigma \to \sigma - \rho/\bo$, the glueball-quarkonium
potential may be eventually expressed in the following form:
\ba
{\cal L}_{h,\sigma} = V_g (h) + \frac{N_F}{48 \pi^2}\, h \left(\sigma -
\ln\frac{ \Lambda_F}{\Lambda_C} - \frac14\right) +
{\cal L}_{\sigma}^{\dim}(S)  - \frac{N_F C_g}{12} \sigma \ , 
\la{gluqu} 
\ea
We see that in \gl{gluqu}
the coupling between glueball and quarkonium fields is fixed unambiguously.

Let us examine the dependence of the QCD vacuum energy on 
slowly varying scalar fields,
$S(x) \simeq const$.
 The quark vacuum energy is connected to
the quark condensate (at 1-loop approach) \ci{No}
\ba
\partial_{\bar S} \, W_F(\bar S)=
 i  \langle \bar \Psi \Psi \rangle  &=&
- \left \langle
\tr { \langle x | \frac{\bar S}{{\slash D}^2 + \bar S^2} \, R | x
\rangle} \right\rangle_{n.p.}
\no
&\to& -\mbox{\rm sign}(\bar S) \,
  \langle \tr{\langle x | \pi \delta({\slash D}^2) \, R | x \rangle}
 \rangle_{n.p.} \qquad ( \bar S \to 0 ) \ ,\\
&\to&
- \frac{1}{\bar S} \, \langle \tr{ \langle x|R|x \rangle} \rangle_{n.p.}
= - \frac{1}{12 \, \bar S} \, C_g
\qquad ( \bar S \to \infty)\ .
\ea
Let us focus our attention on the case $S \gg \lam$.
{}From eq.~\gl{eq10} we have
\be
Z (\rho, S \to \infty)
\to \int \D{G} \exp\left\{ - \int d^4 x \,
 \frac{1}{4} \left(\frac{1}{\tilde g^2}- N_F \,
\frac{1}{24 \pi^2} \,
\ln \frac{S^2}{\mu^2}   \right)\, (G_{\mu\nu}^a)^2
+ O(N_F)^2 \right\}_{n.p.} \ . \la{asympt}
\ee
As in \gl{lam1} one can make the RG improvement by redefining the 
strong coupling constant,
\be
\bar g^{-2}
= \tilde g^{-2} - N_F \, \frac{1}{24\pi^2} \,
\ln \frac{S^2}{\mu^2} \simeq
\frac{11 N_c}{48 \pi^2} \, \ln \frac{\mu^2}{\tilde \lam^2}
 - \frac{2 N_F}{48 \pi^2} \, \ln \frac{S^2}{\tilde \lam^2}
\equiv
\frac{11 N_c}{48\pi^2} \, \ln \frac{\mu^2}{\bar \lam ^2} \ ,
\ee
so that
\be
\bar \lam = \widetilde \lam \, \left( \frac{|S|}{\widetilde \lam} 
\right)^\frac{2N_F}{11 N_c} \ .
\label{barlam}
\ee
Consequently, the leading contribution into vacuum energy
for large $S$ factorizes,
\be
\Evac ( S \to \infty , \rho )
= \Ezero (\rho) \, \left( \frac{|S|}{\lam} \right)^{\frac{8 N_F}{11 N_c}}
+ O(N_F^2) < 0 \ .
\ee
In all cases 
quark polarization effects  lower
the QCD vacuum energy.

Passing on we remark that this result 
is related  to conclusions in ref.~\cite{quigg}.
Indeed let us assume the value of the strong coupling constant
and therefore of  $\lam$
is fixed at a scale much larger than the masses of {\it all} $N_F = 6$
participating
quarks (say e.g.\ a GUT scale $M_U$). Let us denote
this scale by $\hat \lam$. If one likes to know
how this is connected to the low-energy $\lam$ with
$N_F=3$ dynamical quarks, one has to integrate
out the heavy quarks step by step. In this way
one extracts the large mass logarithms in a similar
manner. The result is
\be
\lam = \hat \lam \,
\left( \frac{m_c m_b m_t}{\hat \lam^3} \right)^{2/27}  > \hat\lam;\quad
27 = 11N_c - 2N_F^{light} \ .
\ee

Thus in this Sec. the structure of bosonized
action for a scalar glueball and  its coupling to quark 
degrees of freedom is determined as well as the asymptotic
conditions on the behavior of quark vacuum energy for very
large but constant scalar sources $S$ are formulated.
We would like to stress that this limit belong to the soft
momentum region and it is safe being related to the decoupling
of heavy quarks, i.e.\ one does not expect any phase transition
or strong coupling phenomena when $S \to \infty$.
Hence one should use this conditions to obtain model independent 
constraints on parameters of effective meson lagrangian.

\section{Asymptotic constraints on chiral constants}

 Let us impose the QCD asymptotics \gl{asympt} to be fulfilled 
also in the extended meson lagrangian
\gl{sdim} for
scalar and pseudoscalar fields.  We normalize $a_4$ according to \gl{a4}  
and $a_3$ as in \gl{modif}. First we switch off 
all the external sources but the scalar one,
i.e.\ we retain the scalar field vertices with constants $a_{21}, a_{11}$.

For these unknown parameters, we examine the
asymptotics $S \to \infty$ ($V,A,P=0$,\\
$ U = 1$).  It is
convenient to introduce a new variable $y$ such that
\be
 S \, y := e^{-\sigma} \leftrightarrow \sigma = - \ln S - \ln y \ .
\ee
We then obtain for the effective potential
\be
W(y,S) = - a_4 \, \ln S -  \, a_4 \, \ln y
+ (\frac14 a_4 y^4 + a_3 y^3 +  a_{21} y^2 + a_{11} y ) \, S^4\ . \la{vacen}
\ee
The first term $ (- a_4 \, \ln S)$
already reproduces the leading logarithmic 
behaviour \gl{asympt} of the QCD vacuum energy with the correct coefficient.
Consequently, after integrating out the field $y$ from
the saddle point approximation, the remaining terms
should at most behave like $(const. + O(1/S^2))$.

Thus the term in front of $S^4$ has to be of higher order in $1/S^2$ 
for $y$ taking its saddle point value $y_{min}$
\be
y_{min}\left(\frac14 a_4 \, y_{min}^3 + a_3 \, y_{min}^2 +   a_{21} \, y_{min} + a_{11} \right) = 
O\left(\frac{1}{S^4}\right); \quad y_{min}\not=0 \ . \la{first}
\ee
For $y_{min}$ being the saddle point, the first derivative of
$W(y,S)$ with respect to $y$ has to vanish,
\be
W'(y) =  a_4 \, y_{min}^3 + 3 \, a_3 \, y_{min}^2 + 2 \, a_{21} \,
y_{min} 
+ a_{11} - \frac{a_4}{S^4} \, \frac{1}{y_{min}} = 0 \ . \la{minim}
\ee
Let us expand the last relation in $1/S^2$, with 
$y_{min} = y_0 + \delta/S^2 + O(1/S^4)$. We obtain
\ba
W'(y) &=& a_4 \, {y_0}^3  + 3\,{a_3}\,{y_0}^2 + 2\,{a_{21}}\,{y_0} + {a_{11}}\no 
&& +
   {\delta \,{\left( 
         3\,{a_4}\,{ y_0 }^2  + 6 a_3\,{y_0} + 2 a_{21}\right) }\over {S^2}}
+ O(\frac{1}{S^4})\ . \la{expan}
\ea
Hence one arrives to the second constraint,
\be
 a_4\,{y_0}^3 + 3\,a_3\,{y_0}^2 + 2\,a_{21}\,{y_0} + 
        a_{11} =0 \ , 
\la{second}
\ee
Two constraints make it possible to find the saddle point value $y_0$
and to estimate one of the coefficients, $a_{11}$:
\ba
y_0& =& \frac{- 4 a_3 \pm \sqrt{16 a_3^2 - 12 a_{21} a_4}}{3 a_4}
% =  \frac{- 16 a_3 \pm \sqrt{40 a_3^2 - 16 a'_{21} a_4}}{C_g}
\ ;\no
a_{11} &=&  y_0 \, \frac{8 a_3^2 - 6 a_4 a_{21}}{9 a_4} +
\frac{4 a_3 a_{21}}{9 a _4} \ .
%\no
%&=& \frac{8 a_3^3}{3C_g^2} \left[
%6 A +  17  \pm 
%\left( \frac52 - A\right)^{3/2}\right];\quad
%A \equiv \frac{C_g a'_{21}}{B_0^2 F_0^4} \ . \la{a11}
\ea
In order to obtain information for the chiral lagrangian,
namely the constants $L_5, L_8$, we perform the large
scalar mass reduction $a_{i(j)} \to a'_{i(j)}$, see eq.~\gl{modif}.
and express the chiral constants in terms of
$B_0, F_0, C_g $ and $a_{21}'$.
Together with the universal relations \gl{chm}  
$a'_{11} =  a'_{12} = a'_{13} = a'_{14} = 8 B_0 I_4$,
$\quad I_5 = 0$ one finds
the parametrization of the chiral constant $L_5$
\be
L_5^{\pm} = \frac{B_0^2 F_0^6}{3C_g^2} \left[ \frac{3}{4} A - \frac{29}{16}
\pm    \left( \frac52 - A\right)^{3/2}\right];\quad 
A \equiv \frac{C_g a'_{21}}{B_0^2 F_0^4} \ . \la{L5m}
\ee
The chiral constant $L_8$ is 
supposed to obey \gl{L8} and will be in agreement with
the experimental estimates for the choice of $L_5 \sim 
3 \cdot 10^{-3}$.
Note that the pre-factor in \gl{L5m} already sets the
right scale $\left(B_0^2 F_0^6/3C_g^2\right) \sim 10^{-3}$.
Therefore the remaining numerical factor in parentheses
is expected to be of order 3, which is in fact only
achieved for the choice $L_5^+$. 
It can be achieved for a value of $a_{21}' \sim
-\frac12 F_0^2 < 0$ which just lie inbetween the
predictions of the chiral quark model 
$a'_{21} = F_0^2 \left(1 - (B_0/2M_d)\right) \sim - (2\div3) F_0^2$
and the
chiral bosonization model $a'_{21} = 3/2 \, F_0^2$, respectively,
Thus we see that the asymptotic constraints impose
rigid conditions on the choice of a bosonization model.

The next-to-leading term of the asymptotics \gl{expan} needs a special care.
If $\delta = 0$  then the next term of order $1/S^4$ contributes
to the quark vacuum energy \gl{vacen} only corrections of the same 
order in virtue of \gl{minim}.
Then $O(1/S^2)$ terms
in the effective potential do not appear. 
That would correspond to the identically 
 vanishing dim-6 gluon condensate \ci{No}  $C^g_{6} \sim \langle G^3
\rangle $. 
As there is no reason {\it a priori} 
to neglect the latter one
we obtain the third  constraint for $\delta \neq 0$,
\be
3\,a_4\,{y_0}^2 + 6 a_3\,{y_0} + 2\,a_{21}  = 0 \ . 
\la{third}
\ee
Together with \gl{first}, \gl{second} we thereby have 3 constraints 
for the variables $a_{21},a_{11},y_0$. 
The analysis of these  3 constraints 
shows that their compatibility holds only if 
$a_{21}, a_{11}\not= 0$.
One then has,
\be
y_0 =  \frac{16 B_0 F_0^2}{ C_g} ;\quad
a_{21} =  \frac{16 B_0^2 F_0^4}{ C_g};  \quad
a_{11} = - \frac{256 B_0^3 F_0^6}{3 C_g^2}\ ;
\la{a21}
\ee
which corresponds to
\be
L_5 = L_5^+ = L_5^- = \frac{B_0^2 \, F_0^6}{48 \, C_g^2} \sim
O(10^{-4}) \ ; 
\qquad A = \frac{5}{2} \ .
\ee
However,
the amount of terms in
the polynomial expansion of  \gl{sdim} 
may still not be enough to make a numerically reliable
interpolation, and it will be  necessary to
include terms of $O(S^4)$ into the
scalar lagrangian \gl{sdim}.
(Note that the unnaturally large value of
$a_{11}$ which is the last
coefficient in the polynomial expansion of \gl{sdim}
will presumably reduce then, leading also to more
realistic values for $L_5$.)
Therefore, the third constraint \gl{third}  cannot be taken
too seriously for a given interpolation, and
neglecting the dim-6 condensate corresponding 
to the solution $\delta = 0$
may well be consistent with
the precision of  the expansion of the bosonization model \gl{sdim},

Nevertheless,
we have seen that the asymptotic constraints can be embedded in
the meson sclalar-pseudoscalar lagrangian, 
giving a powerful tool to estimate
some of the chiral coefficients.

Evidently the constants in 
other vertices of the lagrangian \gl{sdim} can be treated in the same manner.
For instance,  
one may expect the mutual  cancellation of large $S$ contributions in
certain combinations
of vertices $a_{22}, a_{12}$ or $a_{23}, a_{14}$,
\be
a_{22} y_{min} +  a_{12} \simeq 0; \qquad a_{23} y_{min} +  2 \,a_{14}
\simeq 0 \ ,
\ee
which however is not anticipated to be accurate 
as only two terms of corresponding expansions are included.

\section{Summary and conclusions}

In this paper 
we have developed the modified bosonization
approach to construct meson lagrangians from 
QCD where the large field but low-energy
asymptotics in QCD is embedded 
after bosonization for scalar field vertices. 

First, it was shown that the chiral bosonization
in different approaches results in the same structure
of vertices (in the large-cutoff approximation),
encoding the information about the QCD dynamics in 
a few chiral constants $I_i$ which can be related to
the phenomenological structure constants $L_i$ and obey
certain model-independent relations among each other.
Next, we have determined the vacuum properties of the
QCD quark energy in the presence of large
external scalar fields and applied the results to estimate
some of the coupling constants in the extended effective
meson lagrangian, including scalar and pseudoscalar fields. 
These ones in turn have been used to obtain an unambigious
estimate of the phenomenologically interesting (but generally rather
model-dependent) structure constants
$L_5$ and $L_8$.

Moreover, by employing conformal bosonization methods
one gets a scalar meson lagrangian which satisfies all QCD
motivated requirements and asymptotics and thereby gives
a direct way to extend it unambigously, such that the
coupling to glueball fields can be described.
As an interesting result, we have given a path-integral
derivation of the effective glueball potential from the
generating functional of QCD using $\delta$-functional
constraints.

There are a few improvements to be done in order to
reach better precision, and one of the most important
is to include higher-dimensional terms into the
polynomial expansion of the lagrangian \gl{sdim}.
In a sense this will lead to an infinite number of
asymptotic sum rules for an infinite number of 
unknown constants, relating them to gluon condensates
of increasing dimension.

Finally, note that in the conformal bosonization approach
dynamical scalar mesons are treated as intrinsic dilatational
modes. It is a challenge to introduce them in the
QCD effective action as true collective variables, in
analogy to chiral rotation modes describing
pseudoscalar mesons.
This program will be realized elsewhere.
\section*{Acknowledgements}
\nn 
One of us (A.A.) is very grateful to Prof. R. Rodenberg for fruitful discussions
and friendly support.

\end{document}